\documentclass[letter,onecolumn]{IEEEtran}
\usepackage{cite}
\usepackage{graphicx}
\usepackage[cmex10]{amsmath}
\usepackage{amssymb}
\usepackage{subfigure}
\usepackage{subfloat} 
\usepackage{graphics}
\usepackage{bigints}
\usepackage{amsthm}
\usepackage{color}
\usepackage{epsfig,psfrag}
\usepackage{epstopdf}
\graphicspath{{Figuras/}}
\usepackage{setspace}

\begin{document}

\title{Outage Probability for Multi-Hop Full-Duplex Decode and Forward MIMO Relay}

\author{Gabriel Fernando Pivaro and Gustavo Fraidenraich. 
\thanks{This work was partially supported by Capes and CNPq, Brazil.} 
\thanks{G. F. Pivaro and G. Fraidenraich are with the Department of Communications, State University of Campinas (Unicamp).}}
\date{\vspace{-5ex}}
\maketitle

\doublespacing

\begin{abstract}
In this paper, a multi-hop (MH) decode-and-forward (DF) multiple-input multiple-output (MIMO) relay network has been studied. To consider a more realistic scenario, Full-Duplex (FD) operation with Relay Self-Interference (RSI) is employed. 

Assuming that the MIMO channels are subject to Rayleigh fading, a simple and compact closed-form outage probability expression has been derived. The key assumption to derive this result is that the mutual information of each channel could be well approximated by a Gaussian random variable. In order to obtain the resultant outage probability, a new excellent accurate approximation has been obtained for the sum of Wishart distributed complex random matrices.

Numerical Monte Carlo simulations have been performed to validate our result. These simulations have shown that, for low and medium interference regime, FD mode performs better than Half-Duplex (HD) mode. On the other hand, when RSI increases, HD mode can outperforms FD mode.
\end{abstract}

\vspace{0.2cm}
\begin {IEEEkeywords}
Multiple-input multiple-output, relay network, Relay Self-Interference, outage probability, Wishart matrices.
\end{IEEEkeywords}

\section{Introduction}

\IEEEPARstart{T}{}he demand for wireless communication technologies in the recent years seems to be unstoppable. The next promising history landmark will be held by 5G networks delivering massive capacity and unforeseen ways of connectivity. In order to achieve higher data rates, one technique that has been considered is Full-Duplex (FD) communication. FD technique increases the spectral efficiency for wireless  systems \cite{Riihonen2011} since it allows devices to transmit and receive simultaneously in the same frequency band. However, in practice such a gain is not attainable as FD nodes suffer from Self-Interference (SI) \cite{Hyungsik2008}, therefore its performance is severely deteriorated.


\textit{Motivation}: Since FD is being considered as a promising technology in conjunction with well established techniques as MIMO and relay, herein, a investigation about multi-hop MIMO relay network has been conducted. In order to assess network performance, the outage probability metric has been used. Moreover, a comparison between FD mode and HD mode has been carried out to establish under what conditions a scheme is more advantageous than the other.

\textit{Prior Related Research}:
The major problem to consider a network where every node is subject to Rayleigh MIMO channel is the mathematical complexity to manipulate random matrices. One way to reduce this complexity is to perform selection/combining techniques such as transmit antenna selection (TAS), space-time block codes (STBC), or maximal-ratio combining (MRC). Those techniques turn the problem more tractable since the final channel gains are characterize by vectors, instead of matrices \cite{Loyka2011}\nocite{Ding2012}\nocite{Kamruzzaman2014}\nocite{Gao2012}\nocite{Ha2013b}\nocite{Sagias2011}\nocite{Fan2012}\nocite{Bao2012}\nocite{Huang2013}$-$\cite{Herath2011}. Here, no selection/combining techniques were employed, therefore, the full capacity of MIMO system is attained.

In \cite{Huang2013}, \cite{Chiani2006}, they have investigated the performance of relay network under co-channel interference and additive white Gaussian noise (AWGN). In \cite{Ding2012} and \cite{Herath2011}, the relay is assumed to be interference-limited, thus AWGN is not considered. In \cite{Annapureddy2011}, if the interference levels are below certain thresholds, they treat interference as noise at the receivers. As stated in \cite{Kim2010}, signal-to-interference-plus-noise ratio (SINR) may not be so great when the interferers also transmit signals at a power level similar to the source. In this case, high signal-to-noise ratio (SNR) approximation, often used in outage analysis, is not enough to provide an accurate outage expression.

\textit{Contribution}: Our main contribution is to derive a simple and compact closed-form expression for the outage probability. This expression is general for any number of antennas and SNR values. In the derivation of this outage probability, we have to deal with linear combination of Wishart random matrices, i.e., $a \mathbf{W}_1+b \mathbf{W}_2$, which is not Wishart distributed, except for the trivial case $a=b$.  In order to circunvent this problem, we propose an extreme accurate approximation based on an upper $U_b$ and lower bounds $L_b$. Another contribution is to show that the mutual information for the MIMO relay with relay Self-interference (RSI) can be well approximated by a Gaussian distribution. This closed-form expression allows the performance evaluation of a MIMO relay network, providing a reliable way to compare FD and HD modes. To the best of our knowledge, no similar results can be found in the literature. 

This article is organized as follows. Section II and Section III present the system model and the outage probability expression, respectively. The performance evaluation is done in Section IV. Finally, Section V brings the work's conclusions.

\vspace{0.1cm}
\section{System Model}

\newcounter{MYtempeqncnt}
\begin{figure*}[!t]
\normalsize
\setcounter{MYtempeqncnt}{\value{equation}}
\centering
\includegraphics[scale=0.42]{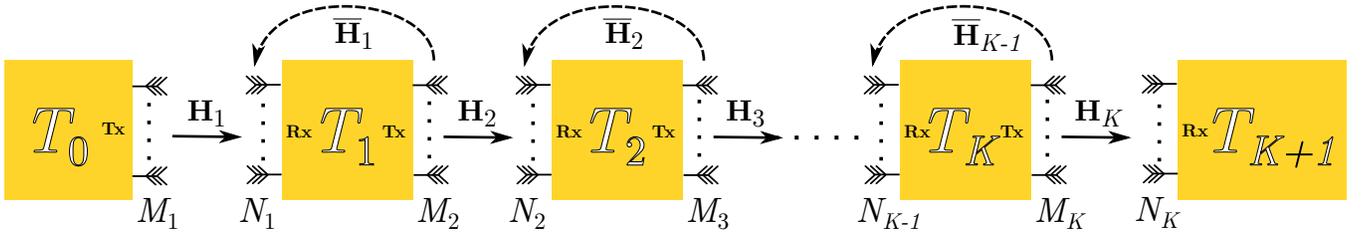}
\caption{System Model of multi-hop MIMO relay network, where the source ($T_0$), equipped with $M_k$ transmitting antennas, communicates with the destination ($T_{K+1}$), equipped with $N_k$ receiving antennas, through $K$ relays, equipped with $M_{k+1}$ and $N_k$ transmitting and receiving antennas, respectively. The solid lines show the desired signal, and the dashed lines show the possible RSI.} 
\label{Channel_model}
\vspace*{4pt}
\end{figure*}


The scenario under consideration is described in \cite{Riihonen2009c}, where all nodes operate in the single-frequency mode and are subject to relay Self-interference. Also, end-to-end communication always consists of multi-hops, i.e., direct communication between the source and the destination is not possible. The system diagram is partially described in \cite{Bao2012} without the feedback link and with addition of RSI. It consists of one source ($T_0$), $K$ relays ($T_K$), and one destination ($T_{K+1}$) as in Fig. \ref{Channel_model}. We also assume that only the last node, $T_{K+1}$, is not subject to RSI, since only reception occurs. The communication protocol used by any relay, $T_k$, is DF. 

Relay Self-Interference (RSI) may occurs when $T_k$ is operating in FD mode, since it transmits and receives simultaneously. The discrete-time received signal $N_k \times 1$ at $T_k$ node from $T_{k-1}$ can be written as \cite{Carleial1983}
\begin{equation}
\mathbf{y}_k=\sqrt{\eta_k}\mathbf{H}_k\mathbf{x}_k+\sqrt{\rho_k}\mathbf{\overline{H}}_k\mathbf{x}_{k+1}+\mathbf{z}_k
\label{eq:received:signal:FD}
\end{equation}
where $\mathbf{x}_k$, $\mathbf{x}_{k+1}$ are $M_k \times 1$, and $M_{k+1} \times 1$ transmitted signals from $T_{k-1}$ and $T_{k}$ nodes. The power constrains on the transmit signals are $\mathbb{E}\footnote{$\mathbb{E[\cdot]}$ denotes the expectation operator.}$[$\mathbf{x}_k^\dagger\mathbf{x}_k$] $\leqslant$ $M_k$. The parameters $\mathbf{\eta}_k$ and $\mathbf{\rho}_k$, are related to the SNR \cite{Marzetta1999}, $\mathbf{\eta}_{k}={SNR}_{k}/{M_k}$, $\mathbf{\rho}_{k}=\overline{{SNR}}_{k+1}/{M_{k+1}}$, where ${{SNR}}_{k}$ and $\overline{SNR}_{k+1}$ are the normalized power ratios of $\mathbf{x}_k$ and $\mathbf{x}_{k+1}$ to the noise at each antenna of $T_k$ node, respectively. Since the last node is not subject to RSI, $\rho_K=0$. The variable $\mathbf{z}_k$ is independent $N_k \times 1$ circularly symmetric complex Gaussian noise vector with distribution $\mathcal{CN}(0, \mathbf{I}_{N_k}$), and uncorrelated to $\mathbf{x}_k$. $\mathbf{I}_{N_k}$ is the identity matrix of order $N_k$. 

The matrix $\mathbf{H}_k$ is $N_k \times M_k$ and represents the channel gain, as depicted in Fig. \ref{Channel_model}. Flat and spatially uncorrelated Rayleigh fading MIMO channels have been considered. So, $\mathbf{H}_k$ is a random and independent matrix. The entries of each matrix are independent and identically distributed (i.i.d.) complex Gaussian variables, with zero-mean, independent real and imaginary parts, each with variance $\sigma^2$. The matrices are known at receiver node only (receiver channel state information-CSIR). The matrix $\mathbf{\overline{H}}_k$ ($k=1,\cdots,K-1$) is a $N_k \times M_{k+1}$ matrix with the same distribution as $\mathbf{H}_k$, and represents the RSI matrix.


Define a Wishart matrix $\mathbf{W}_k$ as \cite{Telatar1999a}
\begin{equation}
\mathbf{W}_k=\left\{\begin{array}{rc}
\mathbf{H}_k\mathbf{H}_k^\dagger\quad N_{k} \leqslant M_k\\
\mathbf{H}_k^\dagger\mathbf{H}_k\quad N_{k} > M_k
\end{array}\right.
\end{equation}
where ``$\dagger$" stands for the conjugate transpose.
Then $\mathbf{W}_k$ is a $m \times m$ with $p$ degrees of freedom random non-negative definite matrix, and thus has real, non-negative eigenvalues, with $p=\max(M_k,N_k)$, $m=\min(M_k,N_k)$. The distribution law of $\mathbf{W}_k$ is called the \textit{uncorrelated central Wishart distribution}, and
\begin{equation}
\mathbf{W}_k\sim\mathcal{W}_m(p,\sigma^2\mathbf{I}_{N_k}).
\label{Wishartdistriution}
\end{equation}

%

When $T_K$ is in HD mode, the transmissions are orthogonal in time and therefore no RSI is generated. In this case, the received signal at $T_k$ node can be written as
\begin{equation}
\mathbf{y}_k=\sqrt{\eta_k}\mathbf{H}_k\mathbf{x}_k+\mathbf{z}_k.
\label{eq:received:signal:HD}
\end{equation}

\vspace{0.1cm}
\section{Outage Probability}
To evaluate performance of a MH MIMO relay system, it's needed to know if the system is able to convey a certain data rate, denoted here by $\mathcal{R}$, and measured in bits/s/Hz. The mutual information ($\mathcal{I}$) gives the dependence between the input and output of the channel, and express the amount of information that can be carried out by the channel. Consequently, the outage probability is the probability that the attempted $\mathcal{I}$ is under certain $\mathcal{R}$, i.e., $\mathcal{P}(\mathcal{R})=\Pr\{\mathcal{I}<\mathcal{R}\}$ \cite{Dohler2010a}.

A MIMO multiple-access channel (MAC) can be assumed since there are two transmitters and one receiver, the first transmission occurs from $T_{k-1}$ to $T_k$ and the second from $T_k$ to $T_k$ (RSI), as can be observed in  \eqref{eq:received:signal:FD} \cite{Carleial1983}. With this assumption, the following rates can be achieved \cite{Tse:2005:FWC:1111206}
\begin{align}
\mathcal{\overline{I}}_{k,fd}&\leqslant \log_2 \left[ \det \left(\mathbf{I}_{N_k} + \rho_k\mathbf{\overline{W}}_{k}\right) \right]\\
\mathcal{I}_{k,fd}+\mathcal{\overline{I}}_{k,fd}&\leqslant \log_2 \left[ \det \left(\mathbf{I}_{N_k} + \eta_k\mathbf{W}_k + \rho_k\mathbf{\overline{W}}_k\right) \right].
\label{RateRR_SR}
\end{align}
where $\mathcal{\overline{I}}_{k,fd}$ denotes the mutual information between transmitter $k$ and its self receiver $k$, and $\mathcal{I}_{k,fd}$ denotes the mutual information between transmitter $k-1$ and receiver $k$.
Using the logarithm properties, and that $\det(\mathbf{AB})=\det(\mathbf{A})\det(\mathbf{B})$ and $\det(\mathbf{A}^{-1})=\det(\mathbf{A})^{-1}$, where $\mathbf{A}$ and $\mathbf{B}$ are square matrices of same dimension, and assuming that the network is operating in one of the corners points of the MIMO MAC capacity region \cite{Tse:2005:FWC:1111206}, then
\begin{equation}
\mathcal{I}_{k,fd}=\log_2\left[\det\left(\mathbf{I}_{N_k}+\frac{\eta_k\mathbf{W}_k}{\rho_k\mathbf{\overline{W}}_k+\mathbf{I}_{N_k}}\right)\right].
\label{MI:sr:FD}
\end{equation}

As a multi-hop case is being treated here, the outage probability in each of the $K$-hops should be evaluated. And the overall outage is determined by the weakest hop. Under the assumption that the hops are subject to independent fading, the outage probability can be written as \cite{Bao2012}
\begin{equation}
\begin{array}{lll}
\mathcal{P}_{FD}(\mathcal{R})&=&1- \Pr \{\mathcal{I}_{1,fd}>\mathcal{R},\dots,\mathcal{I}_{K,fd}>\mathcal{R}\}
\\&=&1-\prod_{k=1}^K [1-\mathcal{P}_{\mathcal{I}_{k,fd}}(\mathcal{R})]
\end{array}
\label{eq:Pout:FD}
\end{equation}
where $\mathcal{P}_{\mathcal{I}_{k,fd}}(\mathcal{R})$ is the cumulative distribution function (CDF) of $\mathcal{I}_{k,fd}$. For the MIMO case, the authors in \cite{Wang2004a} have shown that $\mathcal{I}$ could be well approximated by a random variable with Gaussian distribution. \emph{Our assumption is that the same holds for the quotient of matrices in the case of considering the RSI.} As will be verified in the numerical results section, this approximation proves to be excellent.

For a Gaussian random variable $X$, the following relation is valid $1-\mathcal{P}_{X}(x)=Q_X(x)$, where $Q(\cdot)$ is the Gaussian $Q$-function given in \cite{Craig1991}. Using this in \eqref{eq:Pout:FD}, the closed-form expression for the outage probability can be written as
\begin{equation}
\mathcal{P}_{FD}(\mathcal{R})=1-\prod_{k=1}^{K+1} Q_{\mathcal{I}_{k,fd}}\left(\frac{\mu_{k,fd}-\mathcal{R}}{\sigma_{k,fd}}\right).
\label{eq:poutage2}
\end{equation}

In order to calculate the outage probability in \eqref{eq:poutage2}, it is necessary to determine $\mu_{k,fd}=\mathbb{E}[\mathcal{I}_{k,fd}]$ and $\sigma_k^2=\mathbb{E}\left[\mathcal{I}_{k,fd}^2\right]-\mu_{k,fd}^2$. Note that, $\mathcal{I}_{k,fd}$ in \eqref{MI:sr:FD},  can written as $\mathcal{I}_{k,fd}=\log_2 \left[ \det \left(\mathbf{I}_{N_k} + \eta_k\mathbf{W}_k + \rho_k\mathbf{\overline{W}}_k\right) \right]-\log_2 \left[ \det \left(\mathbf{I}_{N_k} + \rho_k\mathbf{\overline{W}}_k\right) \right]$. To characterize this random variable, the distribution of $\eta_k\mathbf{W}_k + \rho_k\mathbf{\overline{W}}_k$ would be necessary. Unfortunately, the distribution of linear combinations of Wishart random matrices is not known, except for the trivial case, $\eta_k=\rho_k$. 

The problem can be simplified using the result of \cite{Fiedler1971b} that states: \textit{Let $\mathbf{\overline{W}}_k$ and $\mathbf{W}_k$ be Hermitian $m \times m$ matrices with eigenvalues $\alpha_1\geqslant\alpha_2 \geqslant\dots \geqslant\alpha_m$ and $\beta_1\geqslant\beta_2 \geqslant\dots \geqslant\beta_m$, respectively. In particular, if $\alpha_m + \beta_m \geqslant 0$ (which is certainly true since both $\mathbf{\overline{W}}_k$ and $\mathbf{W}_k$ are positive semi definite} \cite{Goodman1963}) \textit{then}
\begin{equation}
L_b \leqslant \det \left(\mathbf{I}_{N_k} + \eta_k\mathbf{W}_k + \rho_k\mathbf{\overline{W}}_k\right)\leqslant U_b
\label{eq:Lower:Upper:Bound:Det}
\end{equation}
where
\begin{equation}
\begin{array}{lll}
L_b=\log_2 \prod_{i=1}^m\left(1+\rho_k\alpha_i+\eta_k\beta_i\right)\\
\\
U_b=\log_2 \prod_{i=1}^m\left(1+\rho_k\alpha_i+\eta_k\beta_{m+1-i}\right)
\end{array}
\label{eq:Lower:Upper:Bound:Det:Proposed}
\end{equation}
\textit{are the lower and upper bounds, respectively}.

%
%

A constant factor 1 has been inserted in \eqref{eq:Lower:Upper:Bound:Det:Proposed} to account for identity matrix present in \eqref{eq:Lower:Upper:Bound:Det}. Also $\eta_k$ and $\rho_k$ have been inserted to account for the signal power at receiving antennas. Since Wishart matrices defined in \eqref{Wishartdistriution} obey \eqref{eq:Lower:Upper:Bound:Det}, the following approximation has been proposed
\begin{equation}
\log_2\left[\det \left(\mathbf{I}_{N_k} + \eta_k\mathbf{W}_k + \rho_k\mathbf{\overline{W}}_k\right)\right]\approx \frac{L_b+U_b}{2}. \label{eq:approx}
\end{equation}

\begin{figure}[tb!]
\centering
\includegraphics[scale=0.95]{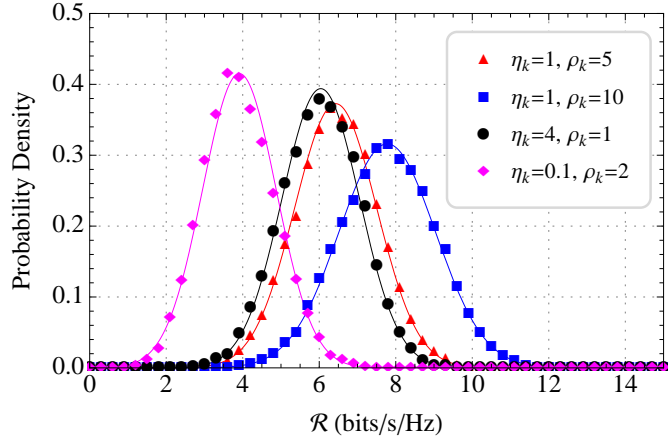}
\caption{Exact (marker points) and approximate (solid line) distribution of $\log_2\left[\det \left(\mathbf{I}_{N_k} + \eta_k\mathbf{W}_k + \rho_k\mathbf{\overline{W}}_k\right)\right]$ for four different sets pairs of ($\eta_k$, $\rho_k$).}
\label{pdf_Approx}
\end{figure}

Fig. \ref{pdf_Approx} illustrates the distribution of $\log_2\left[\det \left(\mathbf{I}_{N_k} + \eta_k\mathbf{W}_k + \rho_k\mathbf{\overline{W}}_k\right)\right]$, considering four different combinations of $\eta_k$ and $\rho_k$. Two important conclusions can be drawn from this figure: 1) the Gaussian approximation is perfectly acceptable; 2) the approximation given in \eqref{eq:approx} proves to be excellent since its difference from the exact distribution is almost indistinguishable.

Therefore, the mean and variance needed are given by
\begin{align}
\mu_{k,fd}&=\mathbb{E}[\mathcal{I}_{k,fd}]\nonumber\\
&=\mathbb{E}\left[\log_2 \left[ \det \left(\mathbf{I}_{N_k} + \eta_k\mathbf{W}_k + \rho_k\mathbf{\overline{W}}_k\right) \right]\right]\nonumber\\
&-\mathbb{E}\left[\log_2 \left[ \det \left(\mathbf{I}_{N_k} + \rho_k\mathbf{\overline{W}}_k\right) \right]\right]\nonumber\\
&\approx \mathbb{E}\left[\frac{L_b+U_b}{2}\right] - \mathbb{E}\left[\Lambda \right]
\label{mean:Ik}
\end{align}

\begin{align}
\sigma_{k,fd}^2&=\mathbb{E}[\mathcal{I}_{k,fd}^2]-\mu_{k,fd}^2\nonumber\\
&\approx \mathbb{E} \left[\left(\frac{L_b+U_b}{2}- \Lambda \right)^2 \right]-\mu_{k,fd}^2
\label{var:Ik}
\end{align}
where
\begin{equation}
\Lambda = \mathbb{E}\left[\log_2 \left[ \det \left(\mathbf{I}_{N_k} + \rho_k\mathbf{\overline{W}}_k\right) \right]\right]=\sum_{i=1}^m  \log_2 \left( 1+\rho_k\alpha_i\right)
\label{eq:Lambda}
\end{equation}
as given in \cite{Telatar1999a}. In order to evaluate the mean $\mathbb{E}[\cdot]$ with respect to the ordered eigenvalues $\alpha_i$ and/or $\beta_i$, the joint probability density function (JPDF) of ordered eigenvalues $\alpha_i$ and $\beta_i$ should be used. The JPDF of ordered eigenvalues $\beta_1\geqslant\beta_2 \geqslant\dots \geqslant\beta_m$ of $\mathbf{W}_k$ is given by \cite{Wang2004a}
\begin{equation}
f(\beta_1,\dots,\beta_m)=\frac{(p-m)!}{m!}\det\left[K(\beta_i,\beta_j)\right]_{i,j=1}^{m} \label{jpdf:betai}
\end{equation}
where
$K(x,y)=\sum_{i=1}^{m} \tilde{\phi}_i(x)  \tilde{\phi}_i(y)$, 
$L_i^d({\beta})=\frac{1}{i!}\mbox{exp}({\beta}){\beta}^{-d} \frac{d}{d{\beta}^i}(e^{-{\beta}}{\beta}^{d+i})$, 
$\tilde{\phi}_i({\beta})=\left[ i!/(i+d)!\right]^\frac{1}{2} L_d^i({\beta}){\beta}^\frac{d}{2}e^\frac{{\beta}}{2}$
and $d={p-m}$. The JPDF of ordered eigenvalues $\alpha_1\geqslant\alpha_2 \geqslant\dots \geqslant\alpha_s$ of $\mathbf{\overline{W}}_k$ is the same as in \eqref{jpdf:betai}.

\section{Numerical Results}
In this section, analysis for the 3-hop case, i.e., $K=2$ has been done. It has been considered that the signal power associated with all hops is the same, and all nodes operate with the same number of antennas. Since the weakest channel determines the capacity, all the three channels have same parameters. 

\begin{figure}[tb!]
\centering
\includegraphics[scale=0.95]{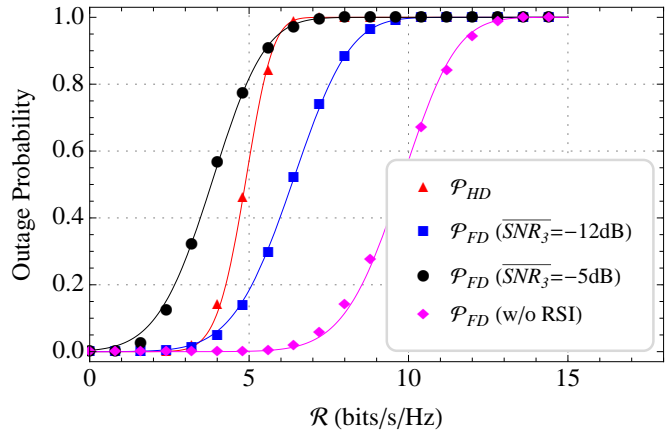}
\caption{Outage probability comparison between full-duplex and half-duplex for the 3-hop network, with $SNR_k=20 \mbox{dB}$, $\overline{SNR_3}=SNR_k-17 \mbox{dB}$, and $M_k=M_{k+1}=N_k=2$, for $k=1,2,3$. The RSI values are $\overline{SNR}_k=SNR_k-5 \mbox{dB}$ and $\overline{SNR}_k=SNR_k-12 \mbox{dB}$. Markers and solid lines show the simulated and analytical values, respectively.}
\label{Outage_FD_HD_DF_2x2x2}
\end{figure}

Fig. \ref{Outage_FD_HD_DF_2x2x2} shows the outage probability for FD and HD modes with $M_k=M_{k+1}=N_k=2$ (MIMO $2\times 2$), for $k=1, 2, 3$, and $SNR_k=20 \mbox{dB}$. For the HD mode, we have used the following expression for the mutual information $\mathcal{I}_{k,hd}=\frac{1}{2}\log_2\left[\det\left(\mathbf{I}_{N_k}+\eta_k\mathbf{W}_k\right)\right]$. 
Notice that the better performance happens when there is no RSI at relays (diamond). The same curve is obtained when the RSI attenuation is grater than $35 \mbox{dB}$. 

Two other RSI scenarios have been considered. One with medium RSI, $12 \mbox{dB}$ attenuation (square). And one with strong RSI, $5 \mbox{dB}$ attenuation (circle). Finally, HD mode scenario is also presented (triangle). It's possible to see that HD mode outperforms FD mode for the last scenario (5 dB attenuation of RSI). 

More importantly, the analytical results and the simulations match perfectly, which validates our Gaussian approximation proposed and also the use of the lower and upper bounds. The numerical simulation was performed with $10^4$ channel realizations.

\section{Conclusion}
In this paper, a simple and compact closed-form outage probability for a $K$ relay network operating in half-duplex and full-duplex modes has been presented. A more realistic scenario, where the relay nodes suffer from RSI was considered. Our assumption is that the mutual information could be well approximated by a Gaussian random variable. Wishart matrices properties have been used to determine the mean and variance of the mutual information.  A comparison between full-duplex and half-duplex was conducted. A perfect agreement between our expression and Monte Carlo simulations results has validated our analysis.






%
%
%

\bibliographystyle{IEEEtran}
\bibliography{library}

\end{document}